\documentclass[aps,prc,twocolumn,showpacs,preprintnumbers,amsmath,amssymb,floatfix]{revtex4-1}
\usepackage{graphicx}
\usepackage{bm}
\usepackage{epsfig}
\usepackage{pstricks}
\usepackage{psboxit}
\usepackage{color}

\begin{document}

\title{Time-Dependent Hartree-Fock Approach to Nuclear Pasta at Finite Temperature}

\author{B. Schuetrumpf$^1$, M. A. Klatt$^2$, K. Iida$^3$, J. A. Maruhn$^1$, K.
Mecke$^2$, P.-G. Reinhard$^2$}

\affiliation{
$^1$Institut fuer Theoretische Physik, 
Universitaet Frankfurt, D-60438 Frankfurt, Germany 
} 
\affiliation{
$^2$Institut fuer Theoretische Physik, Universitaet Erlangen-Nuernberg,
D-91058 Erlangen, Germany
} 
\affiliation{
$^3$Department of Natural Science, Kochi University, 2-5-1 Akebono-cho, Kochi
780-8520, Japan
} 

\date{\today}

\begin{abstract}
We present simulations of neutron-rich matter at subnuclear densities, like
supernova matter, with the time-dependent Hartree-Fock approximation at
temperatures of several MeV. The initial state consists of $\alpha$ particles
randomly distributed in space that have a Maxwell-Boltzmann distribution in
momentum space. Adding a neutron background initialized with Fermi distributed
plane waves the calculations reflect a reasonable approximation of astrophysical
matter. This matter evolves into spherical, rod-like, and slab-like shapes and
mixtures thereof. The simulations employ a full Skyrme interaction in a periodic
three-dimensional grid. By an improved morphological analysis based on Minkowski
functionals, all eight pasta shapes can be uniquely identified by the sign of
only two valuations, namely the Euler characteristic and the integral mean
curvature. In addition, we propose the variance in the cell density distribution
as a measure to distinguish pasta matter from uniform matter. 
\end{abstract}


\maketitle
\section{Introduction\label{sec:intro}}

Core-collapse supernovae (see, \cite{Bethe,Suzuki} for a review) are considered
to be relevant for many astrophysical phenomena including synthesis of heavy
nuclei, acceleration of cosmic rays, and formation of neutron stars and black
holes.  The mechanism of such supernovae, however, has yet to be clarified,
although hydrodynamics simulation studies of stellar collapse and subsequent
explosion continue to be improved with respect to dimensionality and neutrino
transport as well as inclusion of general relativity, rotation, and magnetic
fields \cite{Burrows}.  For such clarification, the properties of matter that
constitutes the supernova cores such as the equation of state, the nuclei
present, and the neutrino opacity are indispensable.  This matter, which is
believed to be charge-neutral and mostly beta-equilibrated while having a
degenerate gas of electron neutrinos trapped inside, is often referred to as
supernova matter.  In early studies of supernova matter below normal nuclear
density and at temperatures of order or lower than 10 MeV, nuclei in equilibrium
were shown to be relatively neutron-rich by using a liquid-drop model
\cite{Lamb}, a Thomas-Fermi theory \cite{Ogasawara}, and a Hartree-Fock (HF)
theory \cite{Bonche}.

At densities just below normal nuclear density, nearly spherical nuclei arranged
in a Coulomb lattice and embedded in a roughly uniform neutralizing background
of electrons are so closely packed that the total surface area becomes very
large.  Then, the system tends to lower the total surface area by elongating the
nuclei and forming nuclear rods \cite{Pethick,Chamel}. It was predicted from
seminal liquid-drop calculations \cite{Ravenhall,Hashimoto} that with further
increase in density, the shapes of inhomogeneous nuclear matter change from rods
to slabs, tubes, and bubbles until the system melts into uniform matter.  These
exotic nuclear configurations are often denoted as pasta nuclei, which arise
from subtle competition between the surface and Coulomb energies. The pasta
nuclei are manifestations of liquid-gas mixed phases of nuclear matter, among
which the phases with rods, slabs, and tubes can be regarded as liquid crystals
\cite{Pethick1998}.

In the conventional liquid-drop approach \cite{Ravenhall,Watanabe2001}, pasta
nuclei were studied by assuming a geometrical shape of nuclear matter and using
the Wigner-Seitz approximation. For more realistic description of pasta nuclei,
multi-dimensional calculations without any assumption of the geometrical shape
were performed in the Thomas-Fermi approach \cite{Williams1985,Lassaut}, the
Hartree-Fock approach \cite{NewtonStone,Magierski,Goegelein}, and the quantum
molecular dynamics (QMD) approach \cite{Maruyama,Sonoda2008}.  In these
calculations, the shapes assumed in the Wigner-Seitz approximation were
reproduced, while some cases did indicate the presence of more complex
structures.

Astrophysically, the possible influence of pasta nuclei on the neutrino opacity
may affect supernova explosions and subsequent protoneutron star cooling.  The
cross section for the neutrino-nucleus scattering depends on the structure of
inhomogeneous nuclear matter \cite{Horowitz2004,Horowitz20042,Sonoda2007}. In
fact, the neutrino scattering processes are no longer coherent in the directions
in which nonspherical nuclei are elongated. This is a great contrast to the case
of roughly spherical nuclei of which the finiteness in any direction leads to
constructive interference in the scattering.

Recently, a time-dependent Hartree-Fock (TDHF) approach has been utilized to
describe pasta nuclei in zero-temperature supernova matter
\cite{Sebille,Sebille2011}.  In contrast to static calculations which
concentrate on the one configuration at minimal energy, the TDHF approach allows
nucleons to explore multiple low-lying configurations in the energy landscape
mapped by the time evolution of the mean field. In this work we perform such
dynamic calculations at finite temperature, which are expected to give a
realistic description of supernova matter as long as two-nucleon collisions can
be ignored.  As we shall see, the resultant map of pasta shapes is basically
consistent with the phase diagrams obtained from the QMD calculations
\cite{Sonoda2008}.

The TDHF simulations produce a great variety of very involved, three dimensional
structures. One needs tools to characterize a given configuration in terms of a
few key numbers. This is simple for single finite nuclei where radius and a
couple of deformation parameters serve very well. However, we encounter here
compounds where much less is known ahead of time. For example, we do not even
know the number of subunits or their connectivity. A very general and powerful
means to quantify complex structures is provided by the Minkowski functionals,
shape descriptors from integral geometry~\cite{Santalo:1976,SchneiderWeil:2008}.
 These measures of structure are important tools in statistical
physics~\cite{Mecke:1998,MeckeStoyan:2000,SchroederTurketal:2010} and pattern
analysis~\cite{LorensenArticle:1987,Mecke:1996}.  They were already introduced
to astronomy for structure analysis of galaxy clusters in
Ref.~\cite{MeckeBuchertWagner:1994} and later used to investigate both point
processes in cosmology~\cite{KerscherMecke:2001,KerscherMeckeSchuecker:2001} and
the cosmic microwave background~\cite{Schmalzing:1999}.  Such Minkowski
functionals provide, of course, the optimal technique to quantify pasta matter
and so classify the different possible shapes. This had already been exploited
in \cite{Watanabe:2003,Watanabe:2004,Sonoda2008}. We aim to continue and deepen
studies along this line.

In Sec. II, the TDHF approach is briefly reviewed. In Sec. III, a
finite-temperature map of pasta shapes for inhomogeneous nuclear matter is
presented. With the help of Minkowski functionals given in Sec. IV,  we do
morphological identifications of various phases in the map of pasta shapes in
Sec. V.  In Sec. VI we investigate thermal fluctuations in the density profiles
and discuss their connection with melting of the inhomogeneous structures. 
Concluding remarks are given in Sec. VII.

\section{TDHF description}

In this framework, we want to describe the pasta structure with the
time-dependent Hartree-Fock (TDHF) approximation.  TDHF is a self-consistent
mean field theory which was originally proposed by Dirac in 1930~\cite{Dirac}.
In general, the TDHF equations are derived from a time-dependent variational
principle with the variational space restricted to one single, time dependent
Slater determinant.  In nuclear applications, TDHF usually employs an effective
interaction as, e.g., the Skyrme force, for a review see~\cite{Bender03}. The
treatment becomes then practically identical to a time-dependent density
functional theory in local density approximation as it is widely used in
electronic systems, see e.g. \cite{Gro96}. Computational limitations restricted
earlier TDHF calculations in system size or symmetry \cite{Neg82aR,Dav85a}. 
With the advance of computational power, it is now possible to perform full 3D
calculations without any symmetry restrictions even for a large number of wave
functions. This led to a revival of nuclear TDHF studies in various dynamical
regimes as, e.g., collective vibrations \cite{Simenel03,Mar05a}, large amplitude
motion in fusion and nuclear reactions \cite{Umar06a,Umar06b}, or, similar as
done here, for simulations of stellar matter  \cite{Sebille,Sebille2011}.  

The TDHF calculations are performed on an equidistant 3D grid in coordinate
space with 16 grid points in each direction with a grid spacing of 1 fm. As
usually done in the physics of extended systems, we simulate infinite matter in
a finite box with periodic boundary conditions. The Coulomb problem is solved
with assumed global neutrality which means that we compensate the proton charges
by a homogeneous negatively charged background which is supposed to simulate the
electron gas around. This amounts to use the Coulomb solver with periodic
boundary conditions and skipping the zero momentum component of the Coulomb
field in Fourier space. The kinetic energy operator is evaluated in Fourier
space and time evolution is performed using a Taylor expansion of the mean-field
propagator. For the present calculations we choose Skyrme parametrization SLy6
which was developed with an emphasis on describing neutron rich
matter~\cite{Chabanat}. The code we use for these calculations goes back to
\cite{Umar1989} and has been used since in many applications to nuclear
resonances, e.g.~\cite{guo2007} and heavy-ion dynamics, e.g.~\cite{loebl2012}.

\section{Map of regimes}
\label{chap:Phadiag}

\subsection{Initialization}
We consider matter with a proton fraction of $X_p=1/3$. In order to produce not
too high excitations, we initialize the calculations by distributing a number
$N_{\alpha}$ of $\alpha$ particles randomly over the grid keeping a minimal
distance between each $\alpha$ core to avoid perturbations by the overlap. To
that end, we use ground wave functions for the $\alpha$ particles produced before
in a stationary HF calculation. The number $N_\alpha$ is varied to produce
systems with different densities. These $\alpha$ cores are distributed in
momentum space following a Maxwell-Boltzmann distribution with a given
initialization temperature $T_\mathrm{init}$. Further neutron states are
introduced as plane wave states using a Fermi distribution with the same
temperature to describe free background neutrons. Note that always both spin up
and spin down states are occupied simultaneously to avoid a large spin
excitation. Finally, before starting the TDHF calculation, these wave functions
have to be ortho-normalized.

This randomized setup leads usually to a system without any symmetry. The time
evolution is assumed to be sufficiently chaotic such that robust final states
properties are reached which depend only on temperature and average density,
although there are some regions where two states compete and the final outcome
depends also on the initialization. The average density is tuned by the number
of particles and the temperature by $T_\mathrm{init}$ in the sampling of
$\alpha$ particles and neutrons. Even with $T_\mathrm{init}=0$ we store a
considerable amount of excitation in the system stemming from the interaction
energy of the $\alpha$ particles.

\subsection{Temperature}

The excitation state of matter is characterized by its actual temperature $T$.
It is, however, difficult to measure temperature in the quantum simulation of a
micro-canonical ensemble as generated by the present TDHF calculations. As a
rough measure, we estimate $T$ by the excitation energy $E^*$ through the
relation $E^*=(\pi^2\,A)(2\varepsilon_\mathrm{F})\,T^2$ where $A$ is the total
number of nucleons and $\varepsilon_\mathrm{F}$ the Fermi energy of the system
\cite{Ash76aB}. The excitation energy $E^*$ is defined as the difference of the
actual energy $E$ to the ground state energy $E_0$. The latter is computed by
solving the static HF problem for the given density. This amounts to
\begin{equation}
  T
  =
  \frac{1}{\pi}\sqrt{2\varepsilon_\mathrm{F}\,\frac{E^*}{N}}
  \, .
\label{eq:temperat}
\end{equation}
This excitation temperature $T$ is to be distinguished from the ``initial
temperature'' $T_\mathrm{init}$ which we use to boost stochastically the initial
ensemble of $\alpha$ particles and the background neutrons. Calculating the
excitation temperature $T$ for the different values of the initial temperature
$T_\mathrm{init}$, we find empirically
$T\approx7\,\mathrm{\,MeV}+\sqrt{T_\mathrm{init}(T_\mathrm{init}+100\,\mathrm{\,
MeV})}/6$. This is taken henceforth for a rough calibration of $T$.

\subsection{First overview}
 
\begin{figure}[t]
\centering
\newsavebox\IBox
\centering
\savebox\IBox{\epsfig{figure=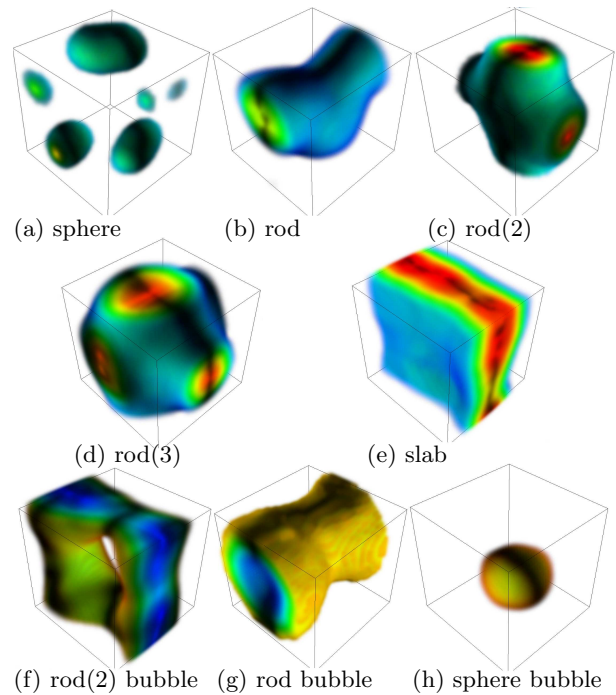,width=8cm}}
 \begin{pspicture}(\wd\IBox,\ht\IBox)
    \rput[lb](-0.,0.){\usebox\IBox}
    \rput[lb](0.,5.8){(a) sphere}
    \rput[lb](2.8,5.8){(b) rod}
    \rput[lb](5.5,5.8){(c) rod(2)}
    \rput[lb](0.8,2.8){(d) rod(3)}
    \rput[lb](4.7,2.8){(e) slab}
    \rput[lb](0.,-0.2){(f) rod(2) bubble}
    \rput[lb](2.7,-0.2){(g) rod bubble}
    \rput[lb](5.3,-0.2){(h) sphere bubble}
  \end{pspicture}
  \caption{\label{fig:pasta}(Color online) Typical shapes of pasta structures at 
    the lowest temperatures.
    These continue to evolve with time but do not change their
    morphological character anymore. Bubble shape illustrations show 
    gas phase, which is indicated by the color-scale (from $0.03 {\rm\, fm}^{-3}$
    (blue/light gray) to $0.12 {\rm\, fm}^{-3}$ (red/dark gray)).} 
\end{figure}

Each setup is evolved in time for 1500 fm/c. After that time, shapes do not
change significantly. The system goes over into a pasta state where some type of
equilibrium is achieved. Depending on the temperature differently strong
fluctuations can be observed.

\begin{figure}[t]
\centering
\newsavebox\PhaBox
\centering
\savebox\PhaBox{\epsfig{figure=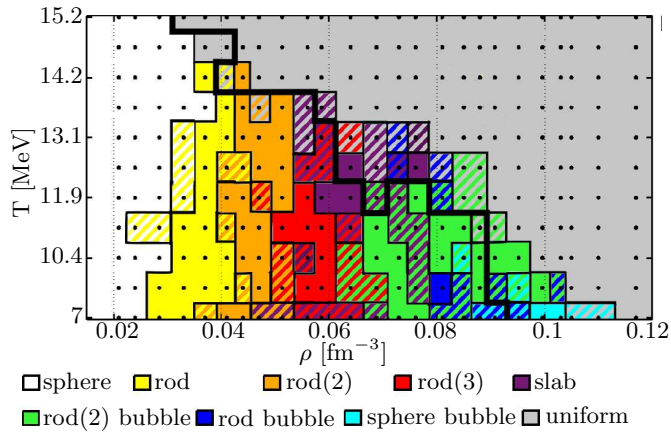,width=8.5cm}}
 \begin{pspicture}(\wd\PhaBox,\ht\PhaBox)
    \rput[lb](0.,0.){\usebox\PhaBox}
    \rput[lb](0.27,0.46){sphere}
    \rput[lb](1.75,0.53){rod}
    \rput[lb](3.5,0.44){rod(2)}
    \rput[lb](5.25,0.44){rod(3)}
    \rput[lb](6.8,0.52){slab}
    \rput[lb](0.27,-0.04){rod(2) bubble}
    \rput[lb](2.6,0.04){rod bubble}
    \rput[lb](4.58,0.0){sphere bubble}
    \rput[lb](6.95,0.07){uniform}
    \rput[rb](0.8,1.4){7}
    \rput[rb](0.8,2.2){10.4}
    \rput[rb](0.8,3.0){11.9}
    \rput[rb](0.8,3.8){13.1}
    \rput[rb](0.8,4.6){14.2}
    \rput[rb](0.8,5.4){15.2}
    \rput[lb]{90}(0.16,2.8){T [MeV]}
    \rput[rb](1.55,1.2){0.02}
    \rput[rb](2.96,1.2){0.04}
    \rput[rb](4.37,1.2){0.06}
    \rput[rb](5.78,1.2){0.08}
    \rput[rb](7.19,1.2){0.1}
    \rput[rb](8.6,1.2){0.12}
    \rput[rb](4.85,0.84){$\rho$ [fm$^{-3}$]}
  \end{pspicture}
\caption{\label{fig:phadiag}(Color online) Map of pasta shapes achieved in
  a TDHF calculation starting from a gas of $\alpha$ particles with
  neutron background for various real temperatures and mean
  densities. The proton fraction is 1/3. Each dot represents two
  calculations. The solid black line shows a phase separation line 
  discussed in Sec.~\ref{chap:Fluctuations}.}
\end{figure}

In FIG.~\ref{fig:pasta} many different pasta shapes are classified. Among the
structures found are rod and slab structures which have been discovered, e.g.,
by QMD calculations~\cite{Sonoda2008}. Rod(2) corresponds to rods forming a
two-dimensional layer. The shape rod(3) describes three rods in x-, y- and
z-direction which cross in one point. Similar shapes were discovered
in~\cite{Sebille,Pais} and the rod(3) structure was also found
in~\cite{NewtonStone}.

FIG.~\ref{fig:phadiag} shows for which density and temperature the different
shapes appear. Note that two calculations for each point in the map were
performed. As we have random initial conditions, the final shapes may differ
although the same values for temperature and mean density are assumed. The
hatched areas with a mixture of colors indicate the different final states
reached in these cases. At higher densities, bubble structures can be seen (the
gas phase has a pasta like shape). Many shapes can coexist here in one point of
the map.

At mid densities, it seems to highly depend on the initial condition wheather
slab or rod(3) shape is formed. For both shapes, the gas phase has the same shapes as
the corresponding liquid phase. So gas and liquid phase symmetrically complement
each other. Hence both shapes are possible for this region.

As mentioned before, we restrict the simulated matter to be periodic by $16
{\rm\, fm}$ in each direction. It is possible that the shapes may differ from
those we would get with other periodic lengths, especially in the mid density
region (rod(3), rod(2) and rod(2) bubble shapes). This has to be checked in future 
work.

\section{Minkowski Functionals}

As already outlined in section \ref{sec:intro}, a powerful tool to quantify the
involved pasta shapes are Minkowski functionals~\cite{MeckeStoyan:2000}. %

There are four Minkowski functionals $W_{\nu}$ defined for a spatial domain $K$
in three dimensions: they are proportional to its volume $W_0 \propto V$, its
surface-area $W_1 \propto A$, the integral mean curvature $W_2 \propto
\int_{\partial K}\mathrm{d}A\, (\kappa_1+\kappa_2)/2$, and the topological
Euler-Poincar\'e characteristic $W_3 \propto \chi$, which is equal to the
integrated Gaussian curvature $\int_{\partial K}\mathrm{d}A\, \kappa_1 \cdot
\kappa_2$. Here, $\kappa_1$ and $\kappa_2$ are the principle curvatures on
$\partial K$, the bounding surface of $K$. The Euler characteristic is a
topological constant. TAB.~\ref{tab:mink_def} summarizes these definitions of
the Minkowski functionals.

Minkowski functionals are useful shape descriptors. They are robust against
noise and have short computation times, both due to their additivity property.
These shape indices form a complete basis of all functionals defined on unions
of convex sets which are translational and rotational invariant, additive, and
at least continuous on convex sets~\cite{Hadwiger:1957}. Thus, the Minkowski
functionals provide a complete scalar morphological characterization of the
pasta shapes these functionals. The Minkowski functionals can be generalized to
Minkowski tensors for an anisotropic structure
characterization~\cite{SchroederTurketal:2010}.

The TDHF results for pasta matter are given as gray-scale data with a density
$\rho_i$ assigned to each voxel $i$ (= grid point).  In order to compute the
Minkowski functionals of a domain $K$, the density field is turned into binary
data via thresholding.  A threshold density $\rho_{th}$ is introduced.  Each
voxel $i$ with $\rho_i\geq \rho_{th}$ is set to black, but if $\rho_i<
\rho_{th}$, white is assigned. The Minkowski functionals of the union of all
black voxels, interpreted as a polygon $K$, can quantify the shape of the pasta
matter in dependence of $\rho_{th}$. Beforehand however, the marching cube
algorithm~\cite{LorensenArticle:1987} is applied, which provides a smoothed
polygonal representation in order to reduce voxelization errors. The Minkowski
functionals of the polygon are evaluated, e.g., using the linear-time algorithm
from Refs.~\cite{SchroederTurketal:2010,SchroederTurketal:arxiv} \footnote{
Implementations can be found at {\tt www.theorie1.physik.fau.de}}.

The threshold density can take on values $\rho_{th} \in
[0,\rho_{\mathrm{max}}]$. But for small thresholds $\rho_{th} \approx 0$ and for
high thresholds $\rho_{th} \approx \rho_{\mathrm{max}}$, the Minkowski
functionals show erratic behavior; thus, the structure values for these
thresholds are ignored~\cite{SchroederTurketal:2010,Mickel2008}.

\begin{table}%
  \caption{\label{tab:mink_def}%
    Minkowski functionals in three dimensions evaluated for a domain
    $K$, with $\kappa_1$ and $\kappa_2$ the principle curvatures on
    $\partial K$.}
  \centering
  \begin{ruledtabular}
    \begin{tabular}{r c l l}
      $W_0$ & $\propto$ & $V$                                                       & volume \\
      $W_1$ & $\propto$ & $\int_{\partial K} \mathrm{d}A $                          & surface \\
      $W_2$ & $\propto$ & $\int_{\partial K} \mathrm{d}A\, (\kappa_1+\kappa_2)/2$   & int. mean curvature \\
      $W_3$ & $\propto$ & $\int_{\partial K} \mathrm{d}A\, \kappa_1 \cdot \kappa_2$ & Euler characteristic \\
    \end{tabular}
  \end{ruledtabular}
\end{table}

\section{Structure classification}

\begin{figure}[t]
\epsfig{figure=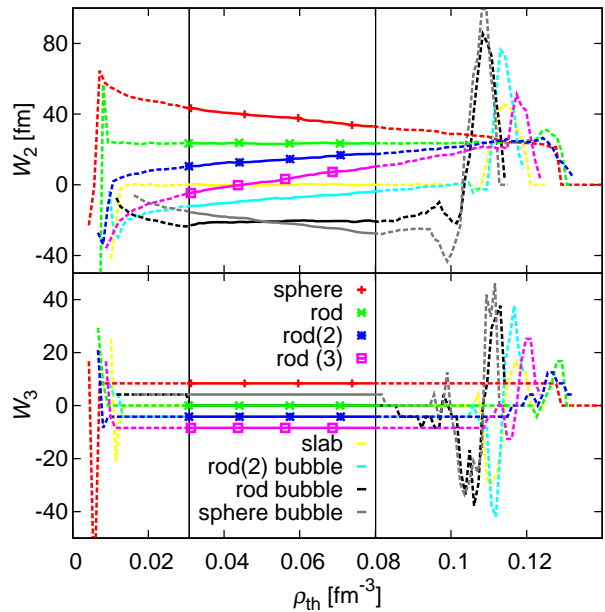,width=8.5cm}
\caption{\label{fig:Mink}(Color online) $W_2$, proportional to the integral mean
  curvature, and $W_3$, proportional to the Euler characteristic as
  function of threshold density for the samples shown in FIG.\ref{fig:pasta}.}
\end{figure}

Calculating the values of the Minkowski scalars for all shapes as a function of
threshold density (FIG.~\ref{fig:Mink}), we see that we can uniquely classify
the different shapes in simple fashion: only the signs of integral mean
curvature and Euler characteristic are needed (see TAB.~\ref{tab:mink}). This
was also found in \cite{Sonoda2008}. However, it is a bit more demanding to
discriminate the rod(3) shape. We need the full trend of the mean curvature
to classify the rod(3) shape correctly. 
\begin{table}[b]%
  \caption{\label{tab:mink}%
    Signs of the integral mean curvature and
    Euler characteristic for each observed shape.}
  \centering
  \begin{ruledtabular}
    \begin{tabular}{c c c c c c c c c}
      shape & sph & rod & rod(2) & rod(3) & slab & rod(2) b & rod b & sph b  \\
      \colrule
      $W_2$ &$>0$ &$>0$ &$>0$    & - to + &$\approx$ 0&$<0$ &$<0$   &$<0$    \\
      $W_3$ &$>0$  &$=0$ &$<0$   &$<0$    &$=0$  &$<0$      &$=0$   &$>0$    \\
    \end{tabular}
  \end{ruledtabular}
\end{table}

The central regime with solid lines in FIG.~\ref{fig:Mink} indicate the physical
values for the threshold density which yield reliably stable values of $W_2$ and
$W_3$ for pasta configurations with the lowest value of temperature are
$\rho_{th}\in[0.03 {\rm\, fm}^{-3},0.08 {\rm\, fm}^{-3}]$. For lower threshold
densities only few dots are recognized as ``white'' (below threshold density)
due to quantum fluctuations in the gas phase. At higher densities quantum
fluctuations of the liquid phase are observed.

Rod(2), rod(2) bubble, and rod(3) consist in channels and tunnels; thus, the
domain has a negative Euler characteristic $\chi < 0$ \cite{Hyde1989}.

The bubble shapes are the complements of the according sphere, rod, or rod(2)
shapes. As the sign of both main curvatures $\kappa_1$ and $\kappa_2$ changes,
the mean curvature $H=(\kappa_1 + \kappa_2)/2$ also changes sign and so does
$W_2$. However, the Gaussian curvature $G=\kappa_1 \cdot \kappa_2$ and thus the
Euler characteristic remain constant.

Both the slab and the rod(3) shape are symmetric in gas and liquid phase, as
described in Sec.~\ref{chap:Phadiag}. The shape of the liquid phase at low
thresholds corresponds to the gas phase at high thresholds. Thus, the integral
mean curvature as a function of the threshold $\rho_{th}$ is point symmetric
w.r.t. a mid value for the threshold density. This is trivially true for the
slab with $W_2 \approx 0$, but holds also for the non-zero mean curvature of
rod(3).

The results indicate that for a complete classification of the observed pasta
shapes, the integral mean curvature as a function of $\rho_{th}$ must be known.

Especially at higher temperatures and mean densities mixtures of phases can
appear, in a sense that at different threshold densities the pasta matter can
take on different shapes (cf. \cite{Sebille2011}). With increasing threshold
densities the shapes can change to shapes which appear usually at lower mean
densities e.g. a rod(2) bubble shape can take on a slab shape at higher values
of $\rho_{th}$. For these cases we take the pasta phase with the bigger range in
$\rho_{th}$.
  
\section{Fluctuations}\label{chap:Fluctuations}

\begin{figure}[t]
\centerline{\epsfig{figure=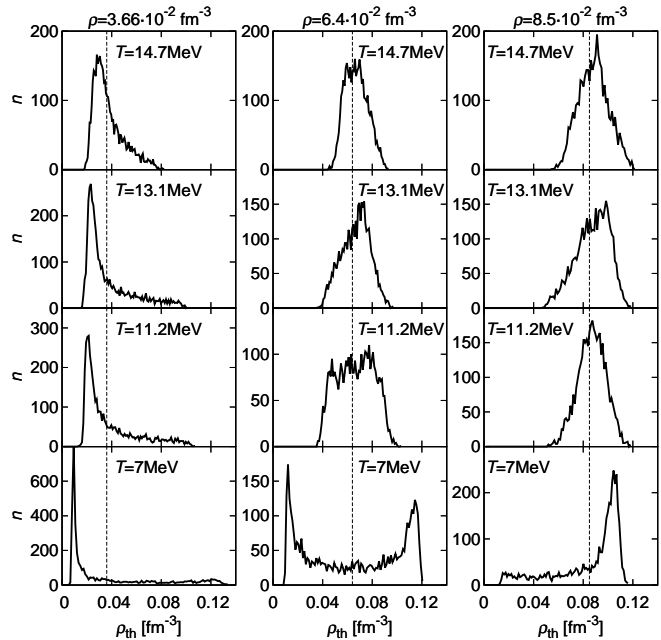,width=9cm}}
\caption{\label{fig:densprof}Threshold density profiles for calculations varying 
        in density and temperature. On each single plot the number 
        of grid points $n$ corresponding to a certain density value is 
        plotted. The dashed lines show the mean density.}
\end{figure}

To show the fraction of liquid and gas phase and the transition to uniform
matter, we plotted the threshold density profiles for various calculations in
FIG.~\ref{fig:densprof}. For the lowest temperature and low mean density, there
is a big peak at small densities and a long tail to high densities. With
increasing mean density a peak at high density develops. At the densities where
rod(3) and slab shapes appear
($5\cdot10^{-2}\mathrm{\,fm^{-3}}<\rho<7\cdot10^{-2}\mathrm{\,fm^{-3}}$) the
peaks for low and high densities have nearly the same width and height. At
higher mean densities the peak for low densities disappears and only a tail
remains.

With increasing temperature the double peak structure vanishes and finally a
single peak around the mean density is forming which moves successively to
central density with ever higher $T$. Consider, e.g., the case of the mean
density of $6.4\cdot10^{-2}\mathrm{\,fm^{-3}}$ in FIG.~\ref{fig:densprof}: We
can observe a double peak structure for the lowest temperature $T=7
\mathrm{\,MeV}$. For $T=11.2 \mathrm{\,MeV}$ we can separate the two phases at a
mid value of the peak and still observe pasta structure. For higher temperatures
no pasta structure can be observed. Only fluctuations around the mean value
plotted as a dashed line remain, so that it is considered as uniform matter.

As the pasta structures vanish with higher density, the variance from the mean
value of the calculation of these plots decreases. Going through the systematics
of the results, we found as a reasonable value for a limit of the variance to
observe pasta structures $2\cdot10^{-4}{\rm \,fm^{-6}}$. The phase separation line
computed with this observable is displayed in FIG.~\ref{fig:phadiag} as a bold
line.

Comparing these results to~\cite{Pais}, the transition density to uniform
matter at $T=7 \mathrm{\,MeV}$ with $9.5\cdot10^{-2}\mathrm{\,fm^{-3}}$ is in
good agreement, but at $T=10 \mathrm{\,MeV}$ we get a higher value of about
$9.0\cdot10^{-2}\mathrm{\,fm^{-3}}$ compared to
$7.7\cdot10^{-2}\mathrm{\,fm^{-3}}$ in \cite{Pais}.

\section{Conclusion}

In this work we have studied the nuclear composition of supernova matter, i.e.
matter at sub-nuclear densities. To this end, we used self-consistent nuclear
mean-field theory at the level of time-dependent Hartree-Fock (TDHF) using the
Skyrme energy functional. We solve the TDHF equations without any symmetry
restrictions on a fully three dimensional coordinate-space grid using periodic
boundary conditions to simulate infinite matter.  The initial state is prepared
by placing $N_\alpha$ $\alpha$ particles stochastically over the simulation box
and adding $2N_\alpha$ neutrons in plane waves which amounts to a 2:1 mix of
neutrons to protons. Scanning cases with different $N_\alpha$ allows to tune
different average densities. These initial states carry already a minimum of
excitation energy because of the interaction energy of the $\alpha$ particles
and the neutrons. We have additionally varied the excitation state by
initializing the $\alpha$ particles and the neutrons with a certain amount of
kinetic energy.  As a measure for the actual excitation, we introduce a rough
estimate for the resulting temperature based on the excitation energy (relative
to the true ground state of the given case). The thus given initial states
evolve to a topologically stable state of pasta matter within about 1000 fm/c. 
This allows us to deduce a map of pasta shapes in the plane of temperature and
density. Those TDHF results agree qualitatively with the phase diagram
calculated in QMD~\cite{Sonoda2008}. In realm of ``rod'' structure, we include
further sub-structures, coined rod(2) and rod(3), which appeared in earlier
Hartree-Fock calculations \cite{NewtonStone,Sebille}. 

As a further means for characterizing these rather involved three-dimensional
structures, we use Minkoswki functionals which have been proven to be useful
measures for structures of fuzzy material. As we deal with smooth density
distributions which cover continuously all values between zero and maximum
density, we produce 0 to 1 step functions by setting a threshold density and we
consider the Minkowski functionals as function of this threshold density.  This
method of analysis allows to characterize the various shapes in terms of a few
key numbers. This allows, e.g., to discriminate reliably the different rod
structures, or to work out the smooth transition to uniform matter at high
temperature. By taking the variance of the density profile as a complementing
further observable, we can define a phase border between uniform matter and
pasta shapes.

It is interesting to study the neutrino cross-section for the pasta shapes
identified here through the Minkowski measures and to check whether the shapes
are associated with corresponding typical pattern in the cross sections. Work in
that direction is in progress.

\begin{acknowledgments}
This work was supported by the Bundesministerium f\"ur Bildung und Forschung
(BMBF) under contract numbers 06ER9063 and 06FY9086, the German science
foundation (DFG) for the grant ME1361/11 as part of the research unit 'Geometry
and Physics of Spatial Random Systems' and by Grants-in-Aid for Scientific
Research on Innovative Areas through No. 24105008 provided by MEXT. The
calculations for this work have been performed on the computer cluster of the
Center for Scientific Computing of J. W. Goethe-Universit\"at Frankfurt. B.S.,
K.I., and J.A.M. acknowledge the hospitality of the Yukawa Institute for
Theoretical Physics, where this work was initiated.
\end{acknowledgments}
\bibliography{Pasta3}

\end{document}